\newcommand{\la}{\langle}
\newcommand{\ra}{\rangle}
\newcommand{\lla}{\la\!\la}
\newcommand{\rra}{\ra\!\ra}
\newcommand{\beq}{\begin{eqnarray}}
\newcommand{\eeq}{\end{eqnarray}}

\newcommand{\eps}{\epsilon}

\newcommand{\btem}{\bibitem}
\newcommand{\TH}{T.\ Hatsuda}
\newcommand{\TK}{T.\ Kunihiro}
\newcommand{\PL}{Phys.\ Lett.\ {\bf B}}
\newcommand{\PTP}{Prog.\ Theor.\ Phys.}
\newcommand{\PR}{Phys.\ Rev.}
\newcommand{\PRL}{Phys.\ Rev. \ Lett.}
\newcommand{\NPB}{Nucl.\ Phys.\ {\bf B}}

\newcommand{\NPA}{Nucl.\ Phys.\ {\bf A}}

\documentstyle[12pt]{article}    

\begin{document}
\begin{flushright}
  RYUTHP 97/3\\
 September 1997\\
\end{flushright}
\begin{center}
{\Large {\bf Some Phenomenological Properties of the Chiral 
 Transition in QCD}}
\footnote{Talk presented at YITP International Workshop on Physics of
 Relativistic Heavy Ion Collisions, 9-11 June 1997, Kyoto, JAPAN.
 To be published in Supplement of Prog. Theor. Phys.}

\vspace{.5cm} 

Teiji Kunihiro\\
\bigskip
Faculty of Science and Technology, Ryukoku University, 
 Seta, Otsu-city\ \ \ \ \ \\ 
 520-21, Japan\\
\end{center}
\begin{abstract}
On the basis of the NJL model as an effective theory of QCD and analogies with
 condensed matter physics, we extract simple physical 
pictures of the chiral phase transition at finite temperature $T$ 
 and/or chemical potential $\mu$ 
 and  predict interesting phenomena associated with the phase
 transition which may be seen in relativistic heavy-ion collisions
 to be performed in RHIC or LHC. We indicate that the macroscopic
 equation describing non-equilibrium phenomena 
 of the chiral transition like DCC might be given  by a 
diffusion-like equation rather than a wave equation 
especially when the baryon density is present.
 
\end{abstract}

\section{Introduction}

The aim of the work to be reported here is to extract simple physical 
pictures of the chiral phase transition at finite temperature $T$ 
 and/or chemical potential $\mu$ 
 and to predict interesting phenomena associated with the phase
 transition which may be seen in relativistic heavy-ion collisions
 to be performed in RHIC or LHC.  The  approach adopted 
 will be not a fundamental 
 nor very microscopic one like lattice QCD reported by 
Kanaya\cite{kanaya}; we shall make semi-phenomenological arguments based 
 on some effective theory of QCD, the NJL model\cite{NJL},
 and also analogies with possible counterparts
 of condensed matter physics.  A part of the present report is based on
a review paper\cite{physrep} based on the NJL model.  A remarkable point 
 is that the basic picture presented in \cite{physrep} has been being 
confirmed in subsequent studies with lattice QCD \cite{lattice} and
 in particular instanton liquid model\cite{instanton}.
  In this report, we shall also 
 present further developments and predictions obtained in the approach 
mentioned above.

\section{The dependence of thermodynamical quantities 
on the number of active flavors and
 the vector coupling}
   
The lattice simulations \cite{kanaya,lattice} show that the order and even the
  existence  of the phase transition(s) are largely dependent on the 
number of the active flavors especially when the physical current 
quark masses  are used: 
For $m_u\sim m_d\sim 10 {\rm MeV} <<100{\rm MeV}\stackrel {<}{\sim}m_s$,
 the phase transition may be weak 1st order or 2nd order or not exist. 

The gross feature of the $T$ dependence and the striking difference 
 between the condensates of u (d) quark and the s quark are
  well described by the NJL model\cite{physrep,HK87,NPB}.
 A minimal model which describes the low energy 
phenomena related with the chiral symmetry may be given
by the  generalized Nambu-Jona-Lasinio (NJL) model 
with the anomaly term\cite{HK3a};
\beq
{\cal L} = \bar{q}(i\gamma \cdot \partial -{\bf m})q 
+ \sum^{8}_{a=0}
{g_{_S} \over 2}[(\bar{q}\lambda_a q)^2 + (\bar{q}i\lambda_a
\gamma_5q)^2] +  g_{_D} [{\rm det}\bar{q}_i (1-\gamma_5) q_j +h.c.],
\eeq
where $\lambda^a$ ($a$=0$\sim$8) are the Gell-Mann matrices with
$\lambda_0$=$\sqrt{2 \over 3}\bf{1}$.
 $SU_V(3)$ symmetry is explicitly broken by the current quark masses,
${\bf m}={\rm diag}(m_u, m_d, m_s)$.
The determinantal term
has the $SU_{_L}(3) \otimes SU_{_R}(3)$  invariance but breaks the  
$U_A(1)$ symmetry;
this term  is a reflection of the axial anomaly in QCD and induced by
 instantons\cite{tH}.
The anomaly term causes a mixing in flavors:
 In the chirally broken phase,
  it induces effective 4-fermion vertices such as
$ \la\bar{d}d\ra (\bar{u}u)(\bar{s}s)$ and
$-\la\bar{d}d\ra (\bar{u}i \gamma_5 u)(\bar{s}i \gamma_5 s)$, where 
 the former (latter) gives rise to 
 a flavor mixing in the scalar (pseudo-scalar) channels.

The numerical calculation of the thermodynamical potential
$\Omega (\lla \bar uu\rra, \lla \bar ss\rra)$ 
 as a function of the quark condensates $\lla \bar {q}_i q_i\rra$ 
 ($i=u, s$) shows the following\cite{unp}:
In the chiral limit, the phase transition is
 of second order, although  the thermodynamical potential is asymmetric 
with respect to the zero condensate due to the cubic term coming from 
the determinantal term. With realistic current quark masses
 ($m_u=m_d=5.5$ MeV, $m_s=135$ MeV), 
$\Omega (\lla \bar uu\rra, 0)$ , i.e., in the $u$-$d$ sector, 
has an asymmetric double well, while $\Omega (0, \lla \bar ss\rra)$, 
i.e., in the strangeness sector, has only single minimum for
 all temperatures, which  implies that the phenomenon of the 
disoriented chiral condensate (DCC)\cite{dcc} could not be expected in the
 strangeness sector in contrast to the non-strange sector.

It is also noteworthy that the entropy $S(T)$ 
calculated with the NJL model 
 shows that $S(T)$ goes up with with increasing $T$ in accordance with
 the fact that the chiral restoration is a phase transition from an
 ordered phase to a disordered phase \cite{NPB}; the increase of the
 entropy may imply \cite{asahat} that the NJL model also qualitatively 
reproduce the ratios $\eps /T^4, (\eps -3P)/T^4$ etc obtained in the 
lattice QCD.
  
 The temperature dependence of the quark condensates shows that at high
 temperatures, the flavor $SU(3)_f$-symmetry becomes worse \cite{NPB}, 
which may reflect in the baryon and the vector meson spectra, because 
 they are well described by the constituent quark models.

So much for the case with $\mu =0$. 
How about the phase transition at finite density?\  A numerical 
calculation\cite{NPB} shows
 that at low temperatures lower than about 50 MeV, the phase transition 
is  of strong first order in the density direction.  Note that  our
 model Lagrangian has no vector term like 
$g_{_V}(\bar{q}\gamma _{\mu}q)^2$.  
 As a matter of fact, the strength and even the existence of the
 1st order transition are strongly dependent on the strength of the 
 vector coupling $g_{_V}$ \cite{asakawa}; the vector term prevents a
  high-density state; see also \S 5.

\section{Mass shift and character change of hadrons}

 The NJL model
\cite{prl,mass}  predicts that  the mass of the sigma meson
 decreases as $T$ is raised till
 $T_c$, a ``critical temperature": $m_{\pi}$ is found to be constant
 as long as $T<T_c$; $T_c$ may be defined as the temperature
 at which $m_{\pi}$ starts to go high. This may be in accordance with
 the lattice results on the screening masses\cite{screening}.
  This suggests that at high temperatures the decay 
$\sigma \rightarrow 2\pi$ would get  suppressed and finally 
hindered, and then only the electro-magnetic process $\sigma \rightarrow
 2\gamma$ is allowed.
  It means that the  sigma meson may show up as a sharp
 resonance with the mass $m_{\sigma}\stackrel {>}{\sim} 2m_{\pi}$.  
Thus we propose to observe $\pi ^{+}\pi^{-},\ 2\pi^0,\ 2\gamma$ and 
 construct the invariant mass and examine whether there is a bump 
 in the mass region 300 to 400 MeV.  
One can expect that in the charged system, the 
 process 
$\sigma \rightarrow \gamma \rightarrow {\rm 2 \ \ leptons}$
become possible, because $\pi ^{+}$ and $\pi ^{-}$ have  different
 chemical potentials with each other\cite{weldon}.

   The behavior of kaon at finite  temperature was 
examined by the present author using the $SU(3)$-NJL model Eq.(1)
\cite{NPB}.
It was found that as long 
as  the system is in the NG phase, the mass of kaon $m_{K}(T)$
 keeps almost a constant, the value   at $T=0$.

One can argue that the vector mesons $\rho, \omega$ and $\phi$
 decrease their masses as the chiral symmetry gets restored; 
 the constituent quark model gives for instance,
 $m_{\phi}\simeq 2M_s$ with $M_s$ being the
 constituent strange quark mass which may be identified with the 
  dynamical one generated by the chiral symmetry breaking\cite{NPB}. 
Other QCD-motivated models and theories also 
predict the decrease of the vector meson masses at finite
$T$ \cite{hatptp}, although it is not necessarily clear 
how the decrease of  the quark condensates is reflected 
in the mass shift in  such theories.

If the strength of the anomaly term decreases, a character change of 
$\eta$ and $\eta '$ mesons is expected. 
 A model calculation\cite{pl88} 
 shows that   the 
$\eta$-$\eta '$ system tends to be well described by 
 the non-strange-strange bases (quark bases) rather than the 
flavor octet-singlet 
ones at high temperatures. 
This has been confirmed also by instanton liquid 
model\cite{instanton}.

\section{Precursory soft modes of chiral transition and channel
 dependence of hadronic correlations at $T>T_c$}

If a phase transition is of second order or weak first order, 
there should exist precursory soft modes in the symmetric 
 phase 
prior to the phase transition to a disordered phase; 
the soft modes become tachyonic on the unstable ``vacuum" after
 the critical point.
This is well known  in condensed matter physics and 
 nuclear physics,  and
 there are lots of examples of such soft modes; see
 \cite{picon} for an example in nuclear physics.
The soft modes are actually fluctuations of the
 order parameter of the phase transition. In the chiral transition,
 thus the soft modes corresponding to the fluctuations  
 $\la\la(\bar \psi \psi)^2\ra\ra$ and  hence 
 $\la\la(\bar \psi i\gamma _5\tau \psi)^2\ra\ra$
due to the chiral symmetry are  expected to exist in the 
 symmetric phase or the high-temperature phase. 
The NJL model was used to demonstrate explicitly that this is the
 case \cite{pl84,prl}.
\footnote{DeTar also conjectured the existence of such modes in a quite
 different context\cite{detar85}.} The subsequent lattice simulations on 
screening masses\cite{screening} and
 studies of the instanton molecules\cite{instanton} 
support the existence of the soft modes.

Remarkably, the lattice simulations \cite{screening,lattice} seemed to 
show that vector-mesonic and baryonic modes were also 
obtained in the high-$T$ phase. Actually, 
the screening masses of the  vector modes coincide
  with $2\pi T$ within the error bars, which may 
simply indicate that the interactions between q-$\bar {\rm q}$'s 
 in this channel are absent or greatly suppressed.
  The nucleons exist as a parity doublet, which  
 was subsequently showed to be compatible with chiral symmetry
\cite{parity}:
A linear sigma model with parity doubling can be constructed which
 is not in contradiction with the low-energy phenomena including the
 $\pi$-N coupling constant.

Is there soft modes in the strangeness sector associated with the
 chiral transition?  Soft modes are expected to exist when the
 phase transition is of second order or weak first order, so that the
 restoring force for fluctuations of the order parameter becomes
 small gradually.  In the
 strangeness  sector, the thermodynamic potential has  single minimum
 and shows an only a shift of the minimum point, which suggests that 
 there is no necessity of a softening of the excitation energy 
 for fluctuations of the order in the strangeness sector. 
A numerical calculation
 shows that this is the case in contrast with the non-strange 
sector\cite{softs}.

\section{Fate of vector couplings and correlations in the
 vector channel at high temperature}

One  may say that there is a reason d'etre of the 
pionic and sigma-mesonic
excitations near the transition point even in the high-$T$ phase; 
they are fluctuations of the order parameter of the chiral 
restoration.
 Then, is there any fundamental reason why  hadronic excitations in the 
 vector channel?  
The approach of the hidden local symmetry\cite{hidden} 
seems 
 to claim that the existence of the vector mesons are intimately related 
 with the chiral symmetry and its spontaneous breaking.   
We show  that 
 the great suppression seen in the screening mass is consistent 
 with the $T$-dependence of the quark-number susceptibility 
$\chi _q(T)$ \cite{qnumber}
 obtained  by  the lattice simulations\cite{qnum2}.

The quark-number  susceptibility $\chi _q$
 is  the measure of the response of the quark
  number density to infinitesimal changes in the chemical potentials 
$\mu _i (i=u,d)$ if we confine ourselves to the two-flavor 
case\cite{qnum2,mcl};
 thus
  \beq
\chi _q(T,\mu)=\frac {\partial} {\partial \mu }
 \rho_q
=\beta \int d{\bf x}\la\la \bar {q}(0, {\bf x})\gamma _0
q(0, {\bf x})\bar {q}(0, {\bf 0})\gamma _0
q(0, {\bf 0})ra\ra,
\eeq
where $\rho _q=\lla 3 \hat {N}_B\rra /V$ is
 the quark-number density with $\hat {N}_B$ being the baryon number 
operator and $\beta=1/T$
 $V$ the volume of the system. 
We remark that the number susceptibility 
 $\chi _q$ at  finite density  is 
 directly related to the (iso-thermal) compressibility
$\kappa _{_{T}}$ as $\kappa_{_T}= {\chi_q}/{\rho ^2}$.
Thus one sees that if $\chi_q$ of a system is large, 
the system is easy to 
 compress, which  may be a reflection of a weak repulsion between 
the constituents of the system. 

The lattice simulations  \cite{qnum2} show that as $T$ is raised 
  $\chi _q$ at $\mu _q=0$ increases
   very rapidly around the critical point of the chiral transition:
 The ratio of the values  in the high-$T$ and low-$T$ phase 
 reads  
${\chi _q(T_c+\epsilon,0)}/{\chi _q(T_c-\epsilon,0)}= 4 \sim 5$
 with the light two flavors, where
 $\epsilon $ is a small number, say $0.03T_c$.

One can easily verify that $\chi_q(T)$ of the free quark gas 
increases  as $M(T)$ decreases
 and reaches $N_fT^2$ at $M(T)=0$: The enhancement is,however, 
 found to be
  merely about 1.6 with $M(T)$ as described in the effective theory
 Eq.(1).
  Thus there must be an additional mechanism to increase  $\chi _q$
 to realize the anomalous (relative) enhancement obtained in the lattice
  QCD. One may first  note that $\chi_q$ is 
the density-density correlation which is
 nothing but the  0-0 component of the vector-vector  correlations or 
fluctuations \cite{mcl}.
Accordingly, $\chi_q$ is further related with
 the retarded Green's function or the response function in the
 vector channel, 
 the poles of which give the masses of the vector mesons. 
Thus one recognizes that the quark-number
 susceptibility is intimately 
related with the properties of the vector mesons and fluctuations 
 in the vector channel.

To demonstrate the relevance of the vector correlations to $\chi_q$,
 a model calculation with a dynamical model was performed \cite{qnumber}
 with a simple NJL model  having a vector coupling.
 Although one only needs  the case with a vanishing chemical 
potential  to compare the results with 
the lattice simulations, it is noteworthy that  when $\mu_q\not=0$
 there arises a  coupling between $\chi_q$ and the scalar-density
  susceptibility $\chi_s$ owing to the non-vanishing ``vector-scalar
 susceptibility" $\chi_{_{VS}}$, which are  defined by
\beq
\chi_s&=&-\frac {d \la\la\bar \psi \psi\ra\ra}{dm}
            =\beta \int d{\bf x}\la\la\bar \psi(0,{\bf x})\psi
             (0,{\bf x})\bar {\psi }(0,{\bf 0})\psi(0,{\bf 0})\ra\ra,
 \nonumber \\  
\chi_{_{VS}}&=& \frac {\partial \la\la\bar {\psi }\psi\ra\ra}
{\partial \mu_q}
            =\beta \int d{\bf x}\la\la
            \bar {\psi}(0,{\bf x})\gamma _0 \psi (0,{\bf x})
\bar {\psi }(0,{\bf 0})\psi(0,{\bf 0})\ra\ra,
\eeq
respectively. One may also note that $\chi _s$ is the fluctuation of 
the order parameter of the chiral transition, and is  related with the 
 sigma meson propagator. Thus  when $\rho_q\not=0$, which
  the lattice QCD is difficult to calculate, the
  properties of the sigma meson reflects in $\chi_q$.
This is a good example of mode-mode coupling known in the dynamic
 critical phenomena in condensed matter physics.
 
We remark that when $\mu _q=0$, the coupling between $\chi _q$ 
and  $\chi _S$ disappear.  Putting  $\mu_q=0$ into the expressions one 
obtains 
$\chi _q={\chi _q^{(0)}(T)}/ {(1+2g_{_V}\chi _q^{(0)}(T))}$, 
where
$\chi _q^{(0)}(T)$ is the susceptibility for the free-quark gas and
 $g_{_V}$ the coupling constant in the vector channel as mentioned in 
 \S 2.
 The denominator of $\chi _q$ is essentially the inverse of the 
 propagator of the vector meson in the ring approximation, 
 at the vanishing four momenta.  One sees that 
$\chi _q$ is suppressed  with the vector coupling; $g_{_V}$ is positive.
It reasonably implies that the system becomes uneasy at $\mu_q\not=0$
  to compress with the vector coupling; recall that $\chi _q$
 is proportional to $\kappa _{_T}$ when $\mu_q\not=0$.

 It was shown  that  the lattice results \cite{qnum2} 
 can be accounted for by 
   a decreasing vector coupling 
as well as the decreasing dynamical quark 
   mass due to the  chiral transition\cite{qnumber}.
Thus the lattice results may simply suggest that the vector 
coupling is  suppressed or vanishes in the hight-$T$ phase, 
 which implies that  
   the vector correlations and hence the collective vector modes would
  disappear in the chirally restored phase.

It is interesting that this picture is consistent with the observation
 that the screening masses of the vector modes obtained in the 
lattice simulations almost coincides with $2\pi T$, the lowest screening
 mass of the q-$\bar {\rm q}$ system in the chiral limit.

\section{What can we learn about DCC from physics of superconductors?}

Recently much attention has been paid to time-dependent 
phenomena of the chiral transition, especially in relation to 
 the disoriented chiral condensate (DCC)\cite{dcc}.
One of the basic questions of this problem is what is a macroscopic
 phenomenological equation which can describe the time-dependent phenomena.
Usually, the phenomenological linear sigma model is used.
This is, however, not necessarily justified.  First we note that 
 the problem is a dynamical critical phenomena of the chiral transition 
 at finite temperature, and there is much resemblance between the chiral
 transition in QCD and the phase transition to a 
superconductors\cite{NJL,physrep}.
 Abraham and Tsuneto\cite{tdgl} examined what type of equation could
 describe slow and long-wave length phenomena in superconductors;
 see also \cite{tin} for subsequent development.
 They clarified that the answer depends on the kinematics in which one 
is interested in:  For example, in the vicinity of the critical point,
 the equation is well approximated by the non-linear {\em diffusion
 equation}, called time-dependent Ginsburg-Landau (TDGL) equation, not
 a hyperbolic equation as derived from the linear sigma model.
 The origin of the dumping is the Landau damping.
 It means that it is an open and hence interesting problem to explore
 what type of equation is suitable to describe the DCC,  and to see
 what kinematical conditions are relevant to relativistic heavy-ion
 collisions. A preliminary calculation shows that when baryon density is
 present, the relaxation effect due to the Landau damping is 
significant\cite{progress}; we notice that the scalar-vector coupling
 is also important at finite $\mu$\cite{qnumber}.

\section{Summary}

We have seen that the static properties, as represented by the quark
 condensates $\lla \bar u u\rra, \ \lla \bar s s\rra$, the 
thermodynamical potential 
$\Omega (\lla \bar u u\rra, \ \lla \bar s s\rra)$,
of the chiral  phase transition 
 can be understood in terms of a simple model Eq.(1).
 We have indicated that the flavor dependence should reflect in the
 phenomena of DCC; DCC may be only relevant to the $u, d$-sector.
 We have also shown that when the baryon density is present, i.e.,
 $\mu \not=0$, the strength of the vector coupling may change the
 nature of the chiral transition substantially.

We have seen that in relation to the mass shift of the sigma meson,
the detection of the  process 
$\sigma \rightarrow 2\pi,\ 2\gamma $ and 2 $\ell ^{\pm}$ is
 interesting.
We have also seen that that $\eta$-$\eta '$ mesons may change
 their characters at high temperatures.

We have also seen that the non-perturbative effects at $T>T_c$ 
are channel dependent:
 the sigma and pionic channel are special because they are intimately 
 related to the chiral symmetry and its spontaneous breaking.
 On the other hand, the interactions in the vector channel are likely 
 to be weak.
It was indicated that the parity doubling in the baryon sector 
is consistent with 
chiral symmetry  and the low energy phenomena at zero temperature.

We have seen that the quark-number susceptibility
 is intimately related to  fluctuations in the vector channel
  in the system: We have suggested that the vector correlations
  seem to be suppressed after the chiral transition.  Direct 
measurements  of this behavior of the susceptibility in experiment
   would be extremely interesting.
We have also emphasized that at finite baryon density, there arises
 a mode-mode coupling between the scalar and the vector channels.

As for DCC, we have indicated on the basis of an analogy with 
 superconductors that the decay process due to the Landau damping is 
 important and it may change the macroscopic equation which describes
 the DCC from the wave-equation like to a diffusion-like equation; 
it also depends on the relevant kinematics.
We remark that in non-equilibrium phenomena like DCC,
 effects of the mode-mode coupling will affect the dynamics 
considerably.

{\bf Acknowledgement}\\ 
 The author thanks Prof. T. Tsuneto for discussions on nonequilibrium 
 phenomena of superconductors and related problems.
    
\end{document}